\documentclass[12pt]{article}

\usepackage{graphicx}
\usepackage{amssymb}
\usepackage{epstopdf}
\usepackage{amsmath,amscd}
\usepackage{color}
\usepackage{graphicx}
\usepackage{latexsym,amssymb,epstopdf}
\usepackage{subfigure}		

\title{Choosing the fittest as a speciation mechanism}
\author{Susanne Schindler, Olaf Breidbach, J\"urgen Jost}
\date{\today}

\newcommand{\barFn}{\bar{F}^n}

\newcommand{\mate}[1]{#1_{\mbox{\tiny mate}}}
\newcommand{\Pm}{\mate{P}}

\newcommand{\Pnt}{P^n_t}

\newcommand{\Ptmign}{P^n_{t, \mbox{\tiny mig}}}

\newcommand{\off}[1]{#1_{\mbox{\tiny off}}}
\newcommand{\Po}{\off{P}}

\newcommand{\bel}[1]{\begin{equation}\label{#1}}

\newcommand{\be}{\begin{equation}}

\newcommand{\ba}{\begin{eqnarray}}
\newcommand{\ea}{\end{eqnarray}}

\newcommand{\qe}{\end{equation}}

\begin{document}
\maketitle

\begin{abstract}
When a population inhabits an inhomogeneous environment, the fitness
value of traits can vary with the position in the environment. Gene
flow caused by random mating can nevertheless prevent that a sexually
reproducing population splits into different species under such
circumstances. This is the problem of sympatric speciation. However,
mating need not be entirely random. Here, we present a model where the
individually advantageous preference for partners of high fitness can
lead to genetic clustering as a precondition for speciation. In
simulations, in appropriate parameter regimes, our model leads to the
rapid fixation of the corresponding alleles.\\

\noindent {\bf Key words:} Speciation, sympatric, sexual selection, mate preferences, fitness-based mating
\end{abstract}

\section{Introduction \label{sec:intro}}

The question how new species arise is central for the theory of biological evolution. When there is no gene flow between subpopulations, different adaptations to the varying circumstances can cause divergent evolution and lead to new species. This is the mechanism of allopatric speciation. The question becomes more difficult for sexually reproducing populations without mating barriers. Here, the homogenizing effect of gene flow can counterbalance the divergent effects of differential selection in a non-homogeneous environment and can prevent the gradual accumulation of genetic and phenotypic differences that are a precondition for speciation. When the habitat is extended, perhaps gene flow is too slow for counteracting the effects of differential selection at the boundaries. This is the mechanism of parapatric speciation. When the habitat is more contiguous, it can still offer different niches that could be utilized by more specialized individuals, but matings between genetically different individuals might always prevent the stabilization of different adaptive specializations within the population.  This is the problem of sympatric speciation. \\
On one hand, it is an empirical question whether such sympatric speciation is possible. Careful case studies seem to have accumulated some evidence for sympatric speciation, both in field studies, e.g., described or surveyed in  \cite{Albertson99,Schliewen01,Sorenson03,Bush04}, the most prominent example being the cichlid fish of West Africa,  and in experiments, e.g., reviewed in \cite{rice_laboratory,kirkpatrick_experiments}. On the other hand, it is a theoretical question to identify mechanisms and to develop formal models that can account for sympatric speciation. It suggests itself to focus on the mating scheme. The simplest assumption is that individuals choose or accept their mating partners randomly within their population. Individuals, however, can potentially increase their reproductive success by being more discriminative and by selecting their mating partners more carefully. This, in turn, will lead to the evolution of traits that make individuals more attractive as mating partners. This is the mechanism 
 of sexual selection identified by Darwin \cite{darwin_origin}. As long as the attractivity of traits is uniform across the population and its habitat, this will not be conducive to the speciation. On the contrary, it will rather produce an additional homogenizing effect because the selection pressure for these traits then is uniform. \\  
When, however, different traits are attractive in different parts of the habitat, this may decrease gene flow and facilitate speciation. The question then is how such differential attractivity can come about. After all, it can only emerge and establish itself when it offers reproductive advantages to individuals.\\
In order to account for sympatric speciation, Dieckmann and Doebeli \cite{Dieckmann99,Dieckmann04a} explored the mechanism of assortative mating which was originally proposed in \cite{MaynardSmith66}. This means that individuals prefer to mate with phenotypically similar individuals and avoid matings with dissimilar members of their population. Doebeli and Dieckmann have demonstrated \cite{Doebeli03} that in the presence of environmental gradients, this can lead to genetic clustering with two (or more) distinct phenotypes inside the population and it can sufficiently reduce the gene flow between these types so that speciation can set in. Two questions arise here. First, how can such a mechanism be implemented? Individuals need to recognize partners that are similar to themselves. In the absence of higher cognitive abilities, this seems to require a genetic linkage between specific phenotypic traits and mating preferences. Second, why is such assortative mating advantageous for individuals? A possible answer is 
 that this might lead to better adapted offspring in a situation where intermediate phenotypes are less fit than more extreme ones. Under appropriate circumstances, assortative mating might be in that way superior to random mating. The question, however, is whether this is the best strategy. \\
The contribution of the present paper then is to propose a simpler
mating strategy with a more basic justification in terms of
fitness, and to demonstrate that this can be a mechanism causing sympatric speciation. The proposed strategy is simply to try to mate with the fittest partner available. This offers the obvious prospect of securing good genes for the offspring. Again, a problem then is how to evaluate the fitness of other individuals and to recognize a fit potential mating partner. This, however, is well studied in the context of sexual selection, and the consequences of such mating schemes have been explored empirically and theoretically. In our context, the question then is how this can lead to phenotypic divergence as a precondition for speciation. In a non-homogeneous environment, phenotypes can have different fitness in different parts of the environment, and in particular, which phenotype is the fittest may vary across the environment. Thus, when in one part of the environment one particular phenotype is the fittest, and in another part another one, then in each part, the fittest phenotypes not only have the advantage of their own superior fitness, but also the additional advantage of access to particularly fit partners when their own fitness makes them preferred. We demonstrate that this double advantage can lead to genetic divergence even in the presence of strong migration between the different parts of the environment. Also, fitness-based mating preferences can lead to assortative mating, thereby providing a more basic evolutionary rationale for the latter. \\ 
A somewhat related model has been proposed by \cite{doorn_origin} who propose a multi-locus computational model of fitness-based mating in which the fitness superiority of males is displayed by a visual cue. Just after the levels of magnitude of male investment into this cue have evolved such that the cue can be used by females as a decision basis, females also evolve stronger preferences for the fittest males. The model of \cite{doorn_origin} differs to ours not only in the incorporation of a third trait, but also in its implementation as a stochastic computer model with larger genomes, quasi-continuouse trait distributions, and complex competition schemes. Our approach presented here is an analytically tractable model in the tradition of classic population-genetic models as those of \cite{felsenstein_rosalia,udovic_ri,odonald_assMating,levene_equilibrium,MaynardSmith66,dickinson_sympatric} or \cite{moore_assMating}. Our approach extends the understanding about divergence inducing mating schemes by proposing fitness-based mating as a simpler mechanism which is profitable for a larger set of fitness landscapes than assortative mating.\\
Fitness-based mating is an advantageous strategy under more general circumstances than assortative mating, however. For this discussion, it is useful to utilize Wright's metaphor of the fitness landscape. Assortative mating is good in a fitness landscape with two peaks, that is, where  extreme phenotypes have a higher fitness than intermediate ones, for instance, where both large and small individuals are fitter than those of intermediate size. In that situation, it is good for a small individual to seek a small mating partner, and analogously for large ones. Fitness-based mating then works when there is some inhomogeneity in the environment, that is, when in some part or niche or under some circumstances, smaller individuals are favored, and larger ones elsewhere. Then in that part, the small individuals are the most desired mating partners. When, however, there is only a single peak in the fitness landscape, that is, when medium size individuals are doing best, then a small one should rather seek a larger partner, and conversely. Thus, in that situation, disassortative mating would be best for those individuals that find themselves away from the fitness peak. This may be biologically unstable, however. Fitness-based mating in that situation would always go for the intermediate types, those that are closest to the fitness peak.  In that way, fitness-based mating, in contrast to assortative mating, automatically adapts to the geometry of the fitness landscape.\\

For the purposes of this article, the term
  ``fitness'' is utilized in a simple and naive manner. We simply
  quantify fitness of an individual as the (expected) contribution to the number of
  offspring as the result of a mating. The contributions of the two
  mates are added to determine the expected number of offspring. In
  particular, the use of the term ``fitness'' is non-reflexive in the
  sense that it does not include the mating strategy. Obviously, since
  in our scheme, 
  the number of offspring does not only depend on an individual, but
  on a mating pair, an individual can increase its number of offspring
  by selecting a good partner, and thereby become ``fitter'' in a
  deeper sense of the term. Hopefully, our naive use of the term ``fitness''
  in the present contribution will not lead to misunderstanding. For a
  conceptual discussion of the fitness concept, we refer to
  \cite{Jost03}.\\\\

Many other mechanisms of sympatric speciation have been proposed and
explored, ranging from sexual conflict \cite{gavrilets_sexualConflict} to  chromosome
rearrangements and other genetic mechanisms \cite{Wolfe03}. As
emphasized in \cite{Mayr82,Dieckmann04}, however, one should
distinguish between evolutionary causes of speciation, like environmental gradients or sexual conflict, and mechanisms operating at the genetic level. 

The starting point of the modern species discussion was Mayr's
\cite{mayr_species} biological species concept. According to this
concept, species are groups of populations that show sexual
reproduction and are reproductively isolated from other such
groups. On this basis, \cite{Breidbach04} developed the  view of a
species as a dynamical balance between the diverging forces of
differential selection in a spatially or ecologically extended and
therefore non-uniform habitat and the converging effects of gene flow
through sexual recombination. (Sympatric) speciation then requires
breaking this dynamical balance. The present contribution provides a
simplified formal model that can be analyzed theoretically and tested
in simulations. 

In fact, there is a large body of literature on mating schemes and speciation,
carefully described and reviewed in the book of Gavrilets
\cite{gavrilets_landscapes}. In order to organize the variety of
schemes proposed in the literature, Gavrilets
\cite{gavrilets_landscapes} (pp.~280--287) developed a general
framework in which non-random mating can be modelled. He proposed
mating pools in which all individuals can potentially mate with each
other. Individuals from different mating pools do not meet and hence
do not mate.  He then distinguished two cases. In the first case,
individuals preferentially join a mating pool and randomly mate within
this pool. In the second case, individuals randomly join the mating
pool and preferentially mate. That means, encounters are random but
matings depend on mate preferences. The fitness-based mating model
falls into the second case where each niche's population forms one
mating pool in which individuals mate preferentially. He also distinguishes similarity-based and matching-based mate
preference. If {\em ``mating is controlled  by a single trait [...] expressed in both sexes''} (e.g., as in assortative mating), he speaks of similarity-based mating. If {\em ``mating is controlled [...] by two different sex-linked traits''}, it is matching-based. However, our scheme of fitness-based mating is neither
similarity- nor matching-based. 
The model is developed and analyzed in detail in the first author's thesis \cite{schindler_fbm2} and in a forthcoming publication of her.

\section{Fitness-based mating \label{sec:fbm}}

\subsection{Biological background}
\label{bioback} 
By Darwin's \cite{darwin_origin} principle of the survival of the fittest, the  mate preference should be  in favour of those  mating candidates that best enhance the fitness prospects of the offspring. Logically, this involves two aspects. One is the fitness of the potential mate itself, and the other one is the compatibility with the own geno- and phenotype. We assume here that within the population that we model, compatibility is not an issue, in the sense that there is no hybrid inferiority  for crossing between members of that population with different geno- or phenotypes. Hybrid inferiority cannot be the starting point of genetic differentiation within a population, but is rather a consequence of that. When we want to understand sympatric speciation, we should rather identify possible causes that trigger such a genetic divergence within a homogeneous population. Therefore, we concentrate on the first aspect, the selection of a mate on the
  basis of evidence for its fitness. 
In fact, many empirical examples of such fitness-based mating have been discovered and studied where individuals apparently include physical, behavioral, or mental properties of potential mates in their mating decisions. This can be seen at elaborated tests prior to mating whose outcome is correlated to the subsequent mating success. Such is found in fighting competitions in harem forming populations (red deer \cite{cluttonBrock_redDeer}, sea lion \cite{fabiani_sealion}, gorillas \cite{kappeler_primates}), lek-mating birds \cite{hoeglund_lek}, territory defending animals like hummingbirds \cite{wolf_hummingBird}, as well as in the traditions of nuptial gifts, for example in balloon flies \cite{kessel_balloon}.
 
In addition, also the handicap hypothesis \cite{zahavi_handicap},
which later developed to the theory of costly signalling, proposes
that exaggerated ornaments or weapons can be seen as signals for
outstanding health and good physical condition because exaggerated
features are hindrances in daily life. For example, the brightness of
coloration can on one hand indicate low parasite affection in fish
\cite{milinski_stickleback} or in birds \cite{hamilton_brightBirds}, but
on the other hand it leads to higher mortality risks
\cite{promislow_costs1}. The darkness and condition of a male lion's mane indicates his testosterone level and hitherto fighting success, but increases also body temperature and abnormal sperm \cite{west_lion}.

In any case, the mechanism of sexual selection discovered by Darwin
\cite{darwin_descent} is based on the transformation of an indicator
of fitness to a direct target of selection. That means, sexual
selection as involved in mate choice no longer operates directly on
survival abilities, but that rather on the indicators of such
abilities. When that happens, these indicators themselves undergo a
selection that need no longer be related to those underlying
abilities, but rather operates by triggering mating attractivity.

\subsection{Preliminary considerations}
In this section, we essentially argue verbally about the effects of
different mating schemes under various circumstances. The arguments
presented, however, are derived from a formal model to
be introduced and discussed below, 
and are supported by computer simulations of that model.\\
We consider a 2-locus haploid model. At each locus, there are two
possible alleles. At the first locus, we can have $A$ or $a$. We
assume that there are
fitness differences  between the carriers of alleles $A$
and $a$, in the sense that matings between $A$-carriers are expected to have more
offspring than those between $a$-carriers, with the number of
offspring for mixed matings in-between. We also assume that the fitness
difference is visible, an issue to be discussed in more detail below. At the second locus, we can
have $M$ or $m$. Carriers of $M$ mate only with carriers of $A$, that
is, with the fittest members of the population, whereas $m$-carriers
mate indiscriminately. We assume here that the difference between $A$
and $a$ can be detected from the phenotype, but the difference between
$M$ and $m$ cannot be seen from the phenotype. (In fact, this is
slightly inconsistent, as $M$ vs.~$m$ leads to a behavioral difference
from
which, in principle, some inference can be made about a certain allele value at the second locus, but we do not grant our creatures sufficient cognitive sophistication
for that.) The question then is under which circumstances allele $M$
is advantageous in comparison to $m$, that is, when does it pay to
forgo mating opportunities in order to get the best mating
partners. The following observations can be readily supported by
formal computations, but those are omitted because they are straightforward. \\
When the probability of matings is proportional to the one of meetings
between the types, except when an $M$-carrier refuses a mating with an
$a$-carrier, then $aM$-types will perform worst, because they not only carry  the
burden of lower fitness, but also have the
disadvantage of finding fewer mating partners. In contrast, $Am$-types  enjoy a higher fitness
themselves, and in addition have the best access to
mating partners. Therefore, $aM$ will go extinct
asymptotically, that is, the $M$-allele will only co-occur with the
$A$-allele. Also, 
either $AM$ or $am$ disappears because both have less mating opportunities than $Am$, except possibly when $Am$ was
initially absent. In
addition, $Am$ and $am$ cannot co-exist, as $Am$ is fitter
than $am$, and since matings between the two are not prevented, $Am$ will
then eventually dominate the population. $AM$ and $Am$ can co-exist,
however, because when only the fitter genotype $A$ is present, the
selection for $M$ disappears. Finally, there exists an unstable
equilibrium with only $AM$ and
$am$ which do not interbreed, and where the equilibrium frequency of $am$ needs to be
correspondingly higher than the one of $AM$ in order to compensate for the lower
fitness.\\
The situation becomes more interesting when we have two niches in one
of which $A$ is fitter whereas in the other one $a$ is more
successful. The phenotypic effect that indicates fitness differences
can arise in two different ways. Either the phenotypes produced by $A$
and $a$ are distinct, and one of them is better in the first, the
other in the second environment. For instance, the camouflage provided
by the coloration patterns can vary between the environments. Or, $A$ leads to a good phenotype in one
environment whereas in the other
environment this phenotype is produced by $a$. For example, different
feeding habits in the different environments may be needed for a well
nourished phenotypic appearance. Both these possibilities will succumb
to the same type of analysis.\\
Thus, an $M$-carrier will attempt to meet with
$A$-carriers in niche 1, and with $a$-carriers in niche 2. 
Without migration, we could then have an equilibrium population
with $AM$ and $Am$-types in niche 1, and with $aM$ and
$am$-types in niche 2. When migration occurs, however, then the $AM$-type in
niche 1 will be less
successful than $Am$,  and analogously the $aM$-type in niche 2 will be less
successful than $am$, because their mating success in the other niche
is lower, and so there will be a higher back-migration of $m$-carriers
than of $M$-carriers. Therefore, the effective
reproduction of $Am$ is higher than of $AM$ in niche 1, and so, the latter type
should become extinct. Analogously, $aM$ should
disappear in niche 2. For the remaining types, $Am$ and $am$, we then simply need
to determine the selection-migration balance. \\
So far, the situation was polygamous, or more precisely, matings were
not costly, and then, obviously, the best strategy is to mate as often
as possible, regardless of the quality of the mates. Nevertheless, the
preceding considerations will aid our thinking below. Also, analogous
considerations apply to assortative mating, that is, when at the
second locus we have allele $B$ vs. $b$, when $AB$ mates
only with $A$, and $aB$ with $a$ only, while $b$-carriers are ready
to accept any partner. Again, in each niche, the first-locus allele
with lower fitness is at a disadvantage, but $b$ profits from
back-migration when competing with $B$.  \\
In any case, for
monogamy the outcome changes (or more generally, when the number of matings is limited). Trivially, if every individual is
assured of finding a mate, then each should mate only with the
fittest. Of course, this is self-contradictory if mating decisions are
reciprocal, because then the less fit individuals will find no
partners willing to accept them. Thus, we should rather assume that an
$M$-carrier mates with an individual of highest fitness if it meets
one, but abstains from other matings, whereas an $m$-carrier mates
with the first agreeing individual it meets (and will then not be
permitted further matings). The outcome will now
depend on both the fitness difference between $A$ and $a$ and on the
original distribution of these two types. When the fitness difference
is large or $A$ is initially sufficiently frequent, $M$ wins out, else
$m$. Again, however, the selection pressure for $M$ decreases when $A$
tends towards dominating the population. \\
The situation becomes more interesting and biologically more realistic
if we introduce genders (female and male) with different mating
strategies. Let us assume that the males mate indiscriminately and try
to achieve as many matings as possible, the biological rationale being
that their mating costs are very low, whereas the females try to mate
only with the fittest available individuals as their mating costs are
high (because of high reproductive investments) and consequently the number of times that they can possibly mate
is strictly limited. (This will then in turn induce fierce competition
between males.) We nevertheless assume autosomal inheritance of
the mate preference allele. Then, it is preferable for
a female to only accept fit mating partners, as long as this does not
substantially decrease her mating opportunities, for more than
one reason. Firstly, she can expect to
derive more offspring from matings with fit partners. Secondly, that offspring can be expected to be
fitter itself; in particular, her male progeny will then be more
desired mating partners for females of subsequent generations. In
turn, a male derives a double benefit from his fitness, as he is not
only fitter himself, but also becomes a more desirable mating
partner. That is, Darwin's sexual selection sets in. And, as already
argued by Darwin, the process can then acquire a dynamics of its
own. It becomes advantageous for a male to produce the phenotype that
is an indicator of genetic fitness, essentially regardless of whether
this is a true indicator of fitness or not, as long as it serves the
purpose of inducing females to accept him as a mate. In turn, for
females, such a partner then becomes desirable, but no longer
primarily because
of his underlying fitness, but because it is advantageous for her to
produce male offspring that inherits the trait for
attractiveness. There are many well documented or at least well argued
examples where this process can go astray, that is, lead to certain
exaggerated features as indicators of sexual attractiveness, but which
are biologically useless, if not detrimental for their
carriers. Therefore, as already mentioned in \ref{bioback}, the concept of costly signalling has been
proposed as a solution to this dilemma. The idea is that while for a
male it is best to cheaply fake an indicator of fitness, for females it is
biologically advantageous  to rely only on signals that are honest
indicators of fitness because they are so costly to produce that they can only be afforded by the
strongest, i.e., the fittest males. As this is well discussed in the
literature, e.g., \cite{johnstone_handicap}, we refrain from presenting examples. \\
We rather analyze the mate preference once more. Assuming that for a
female a preference for fit (or fit looking, as discussed) males pays
off, it then becomes beneficial for a male to also pass that mating
preference on to his female offspring even though in the situation
discussed here it plays no direct role in the male line. Therefore,
the mating preference allele should be passed on autosomally and not
become linked to a sex chromosome. \\
In any case, our purpose here is not to contribute to the theory of
costly or honest signalling. We rather want to identify a simple
mechanism that can trigger speciation in populations in varying
environments. We shall therefore assume that genetic fitness is
correctly signalled by the phenotypic expression of the underlying
allele ($A$ vs.\ $a$). As discussed, this is a simplifying assumption,
but it will allow us to concentrate on a basic mechanism for
sympatric speciation.

\subsection{Model setup}

Returning to population genetics, our basic
model includes two niches between which  in\-di\-vi\-du\-als can
migrate. The two niches are ecologically different in the sense that
in each of them a different genotype has the highest
fitness. Individuals have two genetic loci. One gene determines the
fitness, the other one determines the probability or propensity for fitness-based
mating. Mating fitness-based means choosing one of the fittest
individuals. As an abbreviation, we call an individual that mates
fitness-based a {\it ``fitmater''} (a single word -- to be distinguished
from ``fit mater'', that is, an individual that is itself fit).  

Individuals are haploid. On one hand, the model could be readily extended to the
diploid case, but on the other hand, that would not lead to new
phenomena. Genetic loci are diallelic. The alleles of the first locus
which determines the fitness value are denoted by $A$ and $a$, and the
alleles of the second locus which determine the mating behavior are denotyed by $M$ and
$m$. Inheritance follows Mendelian rules.  The allele $A$ is
advantageous in one niche, and $a$ is better in the other niche. While
$m$-individuals mate randomly, 
carriers of $M$ are fitmaters with probability $\mu \in
(0,1]$. Pure fitness-based mating then corresponds to $\mu=1$, and we
shall mostly consider this case only.  In principle, however, it is
useful to have such a parameter available, because one can then
differentiate suitable quantities like the expected number of
offspring w.r.t.\,this parameter. In fact, some of our underlying
analysis has been carried out in such a manner.

Let us first analyze the situation for a single niche where $A$ is
fitter, in the sense that there is a parameter $f>0$ which translates
into fitness values for pairs according to the rules that a pair of
two $a$-carriers has the value $1$, a pair of
two $A$-carriers the value $1+f$, while a mixed pair gets
$1+\frac{f}{2}$. The expected number of offspring produced by such a
pair then is proportional to that fitness value, where the
proportionality factor 
may be chosen such that the total population size stays constant over
time. As is typical for such models, we assume that the generations do
not overlap, that is, the members of each generation are born at
the same time, mate, produce offspring representing the next
generation, and then die. Females mate once with a single male that
the females can either choose randomly or select on the basis of his fitness
value. Males can mate as often as they are accepted by a female,
regardless of her fitness value. 

It is now straighforward to analyze the effect of an increase of
$\mu$. Let us assume that a female switches from random to fitmating
and exchanges an $a$-partner against an $A$-carrier. We want
to compute the effects $\delta p(A), \delta p(a)$ of the switch, i.e., the frequency change caused by the switch of her $A$ ($a$)-offspring. The female is
an $A$ ($a$)-carrier herself with probability $p(A)$ ($p(a)$). If
she has $a$, then she had produced 1 offspring of type $a$ with the
$a$-male, but now she is expected to produce
$\frac{1}{2}(1+\frac{f}{2})$ offspring of each type $A$ and $a$ with
the $A$-partner. Likewise, if she has $A$, then she had produced
$\frac{1}{2}(1+\frac{f}{2})$ offspring of each type $A$ and $a$ with the
$a$-partner, but now she is expected to produce $1+f$ $A$-offspring
with her $A$-mate. Thus,
\ba
\label{1}
\delta p(a)&=& \frac{1}{2} \left(\frac{f}{2}-1\right)p(a)
-\frac{1}{2}\left(\frac{f}{2}+1\right)p(A)\\
\delta p(A)&=& \frac{1}{2} \left(\frac{f}{2}+1\right)p(a)
+\frac{1}{2}\left(\frac{3f}{2}+1\right)p(A).
\ea
From this, one readily checks that
\be
\frac{p(a)+\delta p(a)}{p(A)+\delta p(A)} =\frac{1 - 2 p(A)}{1 + 2 p(A)} < \frac{p(a)}{p(A)},
\qe
that is, this leads to a decrease of the number of
$a$-carriers. Actually, this may seem obvious, but it is not entirely
trivial because the expected increase in reproductive success of
$a$-females may be relatively stronger than that of $A$-females
when they are in the minority, $f$ is large, and fitmating is
prevalent. Nevertheless, this does not lead to an increase of $a$ in
the population because after switching, the female $a$-carriers are no longer
breeding true, and the male $a$-carriers loose their mates. Thus,
fitmating will make the selective advantage of $A$-carriers even
stronger. 

It is also clear that fitmating is a superior strategy in terms of the
expected number of offspring for less fit females than assortative
mating whereas it does not make a difference for fit females. For less
fit females, assortative mating is not a good option because that
would require them to choose equally less fit mates. In particular, as
long as  a subpopulation of $a$-individuals persists, we do not
expect an allele for assortative mating to become fixed in the
population, in contrast to our fitmating allele $M$.

When individuals can now migrate into the niche under consideration
and mate there, coming
from another niche where $a$ is fitter than $A$ and where therefore
the $a$-carriers are in the majority, then this will induce a decrease
of the frequency of $A$ in our niche. When there are fewer
$A$-carriers around, however, then fitmating becomes more
advantageous, simply because then the chances are higher to randomly
encounter an $a$-male and then to switch
from that less fit $a$-carrier to a fitter $A$-male. (Or putting it
the other way around,
if most of the males are $A$ anyway, then chances are that already a
random choice will lead to an $A$-partner, and therefore, there is
little expected gain from trying to be selective.)  Thus, migration
increases the selective pressure for fitmating. Consequently, since we
have already shown that fitmating in turn increases the selective
advantage of $A$-carriers, we overall see a stronger countereffect to
the immigration of less fit $a$-carriers. In the other niche, in
contrast, the same effect works in favor of  $a$, as it is assumed to
be the fitter one there. This is our rationale for
proposing fitmating as a possible mechanism for inducing speciation in
populations occupying niches with different selective pressures.

% The parameter $p_m$ denotes the migration rate by which individuals
% move from one niche to the other before mating. 

% After migration, each individual chooses a partner, mates and remains in the mating pool. Therefore the population is polygamous, but it is assured that each individual mates at least once. This contrasts natural polygamous populations in which a relatively small number of individuals reproduce relatively often at the expense of many non reproducing individuals. In this way, our model allows for more genetic mixing than in natural populations. In other words, we artificially introduce circumstances that make genetic divergence more difficult. Thus, if our model leads to speciation, then this should be even easier in natural settings. 

% The number of offspring of a couple is determined additively from paren\-tal fitness values. Thus, an individual profits from a mate with high fitness, because it receives more offspring by its mate's contribution. \\

We now come to the dynamics produced by our model. 
In order to state  the recurrence equation for the  change in time of
the frequency of each genotype combination in the subsequent
generation, we need some  notation: $t\in\mathbb{N}_0$ denotes the
current generation. The parameter $\alpha \in \{A,a\}$ stands for the allele at the
first locus, $\nu \in \{M,m\}$ for the one at the second
locus. The frequency of individuals with
genotype $(\alpha,\nu)$ in niche $n$ at time
$t$ is denoted by $p^n_t(\alpha,\nu)$. This frequency after individuals have
had the opportunity to migrate is denoted by $\Pnt(\alpha,\nu)$. The probability that offspring of type $(\gamma,\omega)$ is produced, given that the parents have the genotypes $(\beta,\rho)$ and $(\alpha,\nu)$, respectively, is denoted by  $\Po(\gamma,\omega|\beta, \rho
;\alpha,\nu)$; the Mendelian inheritance rules determine $\Po$. The
probability that $(\alpha,\nu)$ chooses $(\beta,\rho)$  for mating is
denoted by $\Pm^n(\beta, \rho |\alpha, \nu )$; it can be computed within our model, see \cite{schindler_fbm2}. The additive fitness value of two parents $\alpha$ and $\beta$ is denoted by $F^n_{\alpha,\beta}$ and normalized by its mean value $\barFn$. The recurrence equation then is (see \cite{schindler_fbm2} for
  the derivation)

\begin{equation}
\label{eq:replicator}
p^n_{t+1}(\gamma,\omega) = \sum_{\alpha,\nu} \Pnt(\alpha,\nu) \sum_{\beta,\rho}\Po(\gamma,\omega|\beta,\rho;\alpha,\nu) \Pm^n(\beta,\rho|\alpha,\nu) \frac{F^n_{\alpha,\beta}}{\barFn}.
\end{equation}

Equation (\ref{eq:replicator}) describes the genotype distribution of the subsequent generation $p^n_{t+1}$, given the current distribution $\Ptmign$ after migration in niche $n$. The equilibrium distribution is then obtained by iterating equation (\ref{eq:replicator}), but also explicit equilibrium solutions have been derived in \cite{schindler_fbm2}. The next section will outline the main features of the model dynamics.

% \subsection{Model analysis}
% We outline two different approaches for the preliminary analysis of the model.
% \begin{enumerate}
% \item Compare the effective fitness, that is, the expected number of
%   offspring produced, of the carriers of different alleles and check
%   whether the $M$-carriers perform better than the $m$-carriers.
% \item Introduce a real parameter $m_s, s\in [0,1]$, for the mate
%   preference, with $m_0$ corresponding to $m$ and $m_1$ to $M$ and
%   determine whether or under which circumstances reproductive success
%   is an increasing or a decreasing function of $s$.
% \end{enumerate}

\subsection{Model behavior}

\begin{figure}
%~/Forschung/Scritto/BreidbachJostSchindler/Visualisierung/makeFigure_auswertungSims_zeitlPlot{ ,_ass}.m
\subfigure[]{\includegraphics[width=7cm]{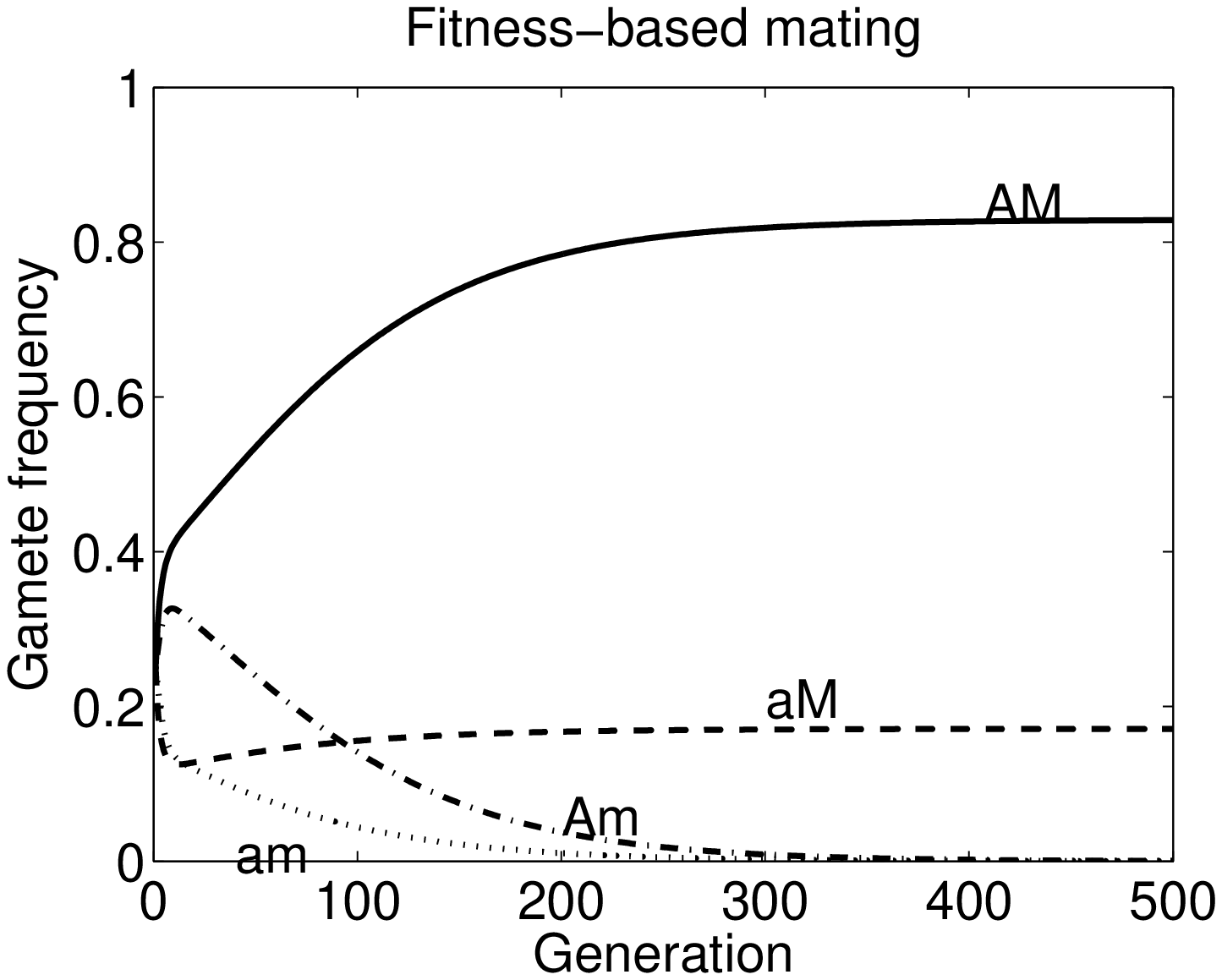}}
\subfigure[]{\includegraphics[width=7cm]{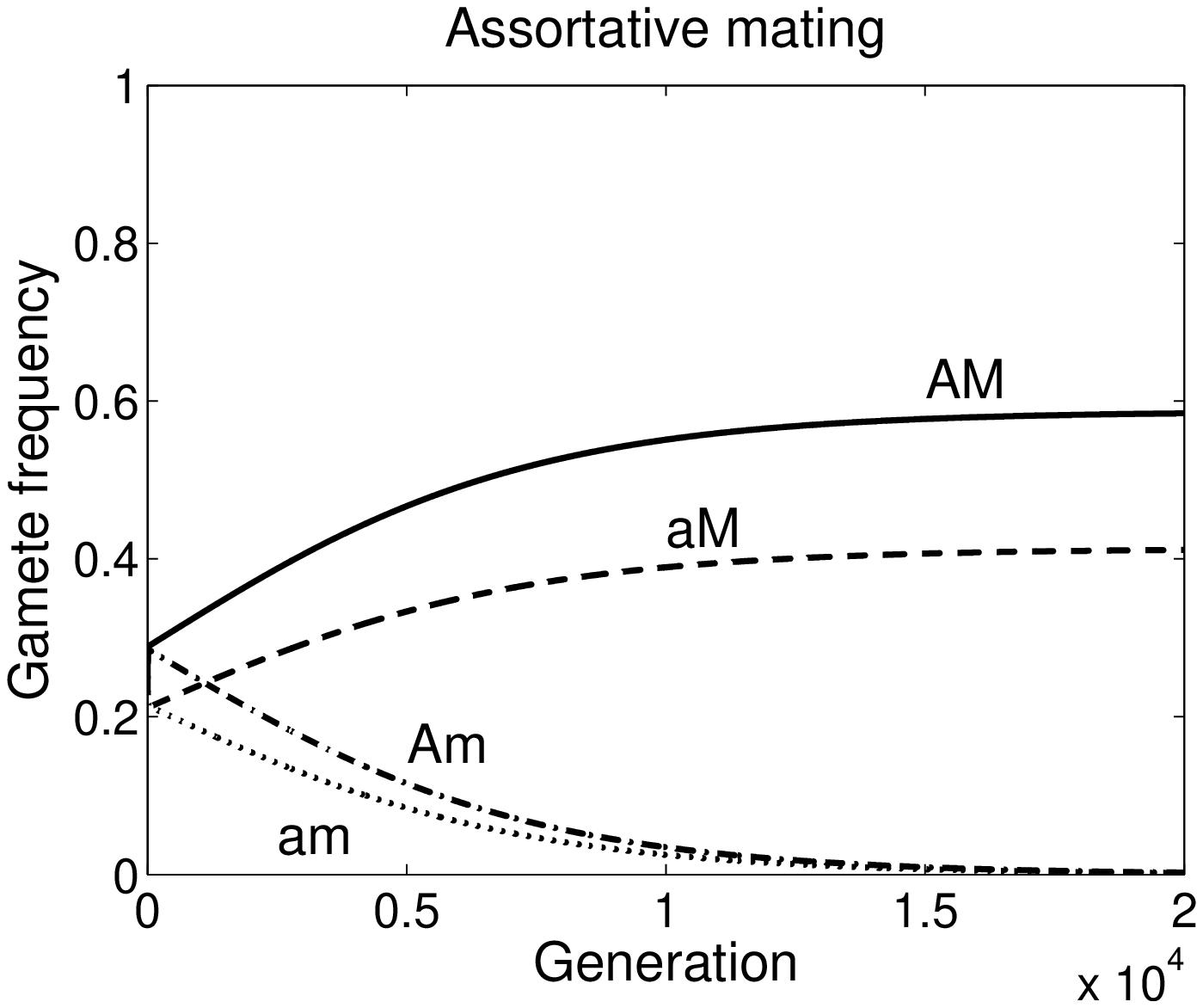}}
\caption{\label{fig:zeitlicheEvo}The initial gamete distribution has been set to a uniform distribution, which would be the result of a randomly mating population where each allele is present at the same frequency. The following parameters have been used: The fraction of individuals migrating from their niche of birth to the other niche is $p_m=0.1$. The parameter $f$ is set to $f=0.1$. The plots show the population development in the first niche, where the allele $A$ is fittest. These distributions are obtained from iterating equation (\ref{eq:replicator}). The situation in niche 2 leads to equivalent results, because the niche conditions are symmetric. (a) Fitness-based mating ($\mu=0.5$) The $m$-carriers mate randomly, and $M$-carriers practise fitmating with a probability of $\mu$ and mate randomly with a probability of $1-\mu$. (b) Assortative mating ($\mu=0.5$) The $m$-carriers mate randomly, and $M$-carriers practise assortative mating with a probability of $\mu$ and mate randomly with a probability of $1-\mu$. The axis ranges up to 20000 generations which compresses the graph at low generation values in a way that the population seems not to start from a uniform distribution, but in fact it does.}
\end{figure}

We now  present some  simulation results
for two different mating schemes in one niche from the iteration of
the recurrence equation (\ref{eq:replicator}) derived from the model. Figure \ref{fig:zeitlicheEvo} shows the population development for fitmating and assortative mating.

\begin{figure}
% ~/Forschung/MatlabFiles/Scritto/BreidbachJostSchindler/makeFigure_auswertungSims_migRate{,_ass}.m
\subfigure[]{\includegraphics[width=7cm]{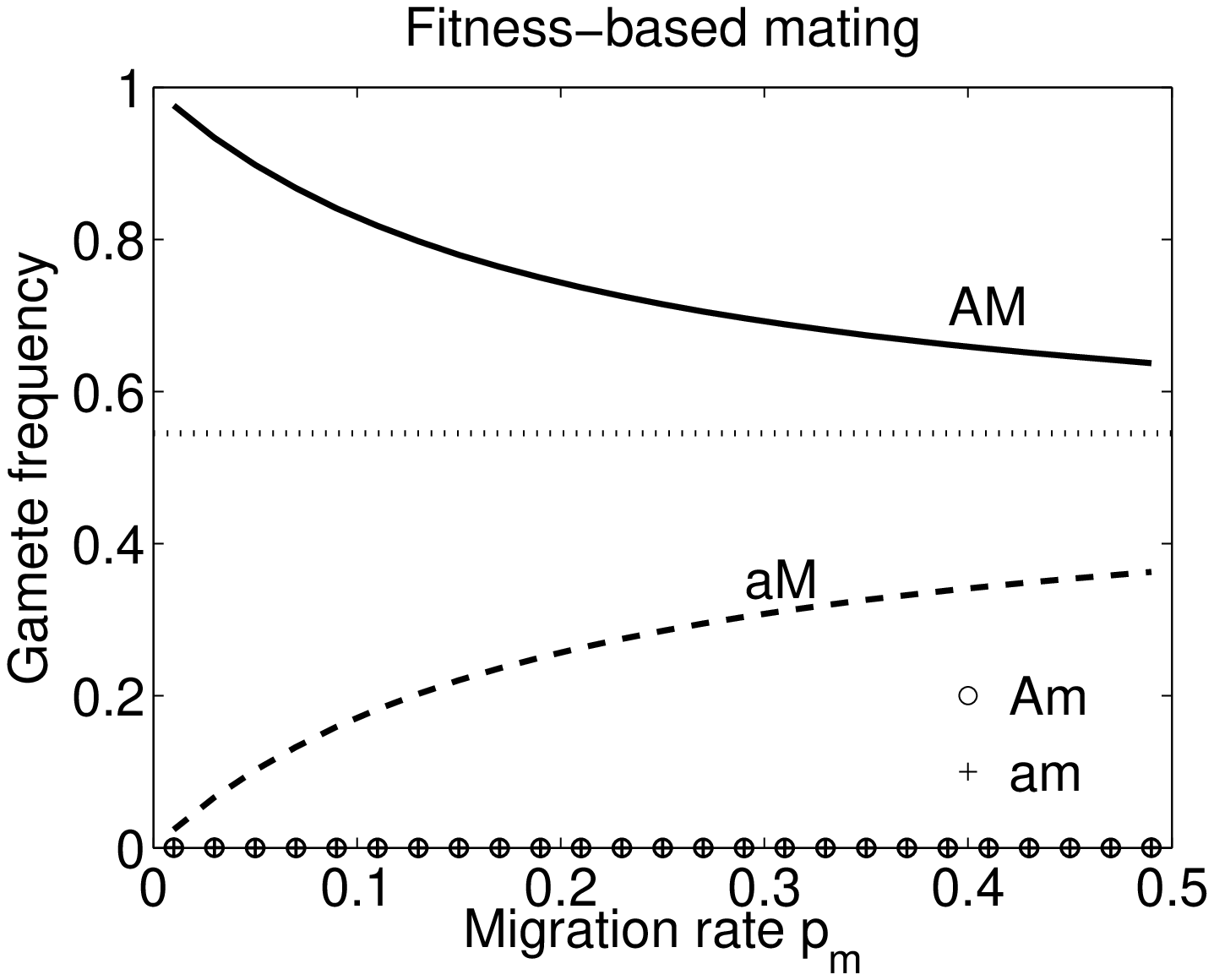}}
\subfigure[]{\includegraphics[width=7cm]{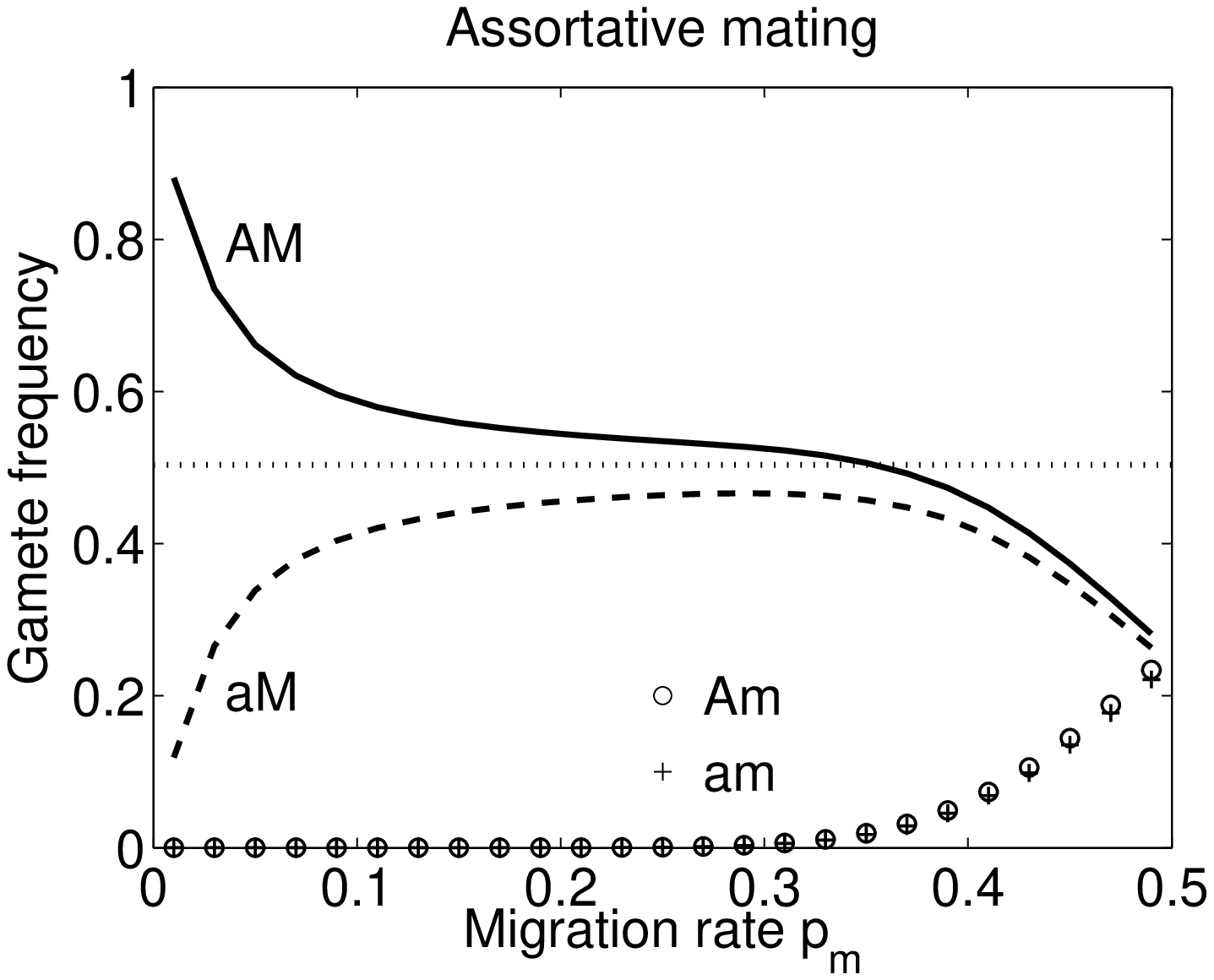}}
\caption{\label{fig:MigRate}The parameters and the initial gamete distribution have been set as in figure \ref{fig:zeitlicheEvo}. The plots show the equilibrium population for various migration rates in the first niche, where the allele $A$ is fittest. These equilibria are obtained from iterating equation (\ref{eq:replicator}). (a) Fitness-based mating ($\mu=0.5$) The $m$-carriers mate randomly, and $M$-carriers practise fitmating with a probability of $\mu$ and mate randomly with a probability of $1-\mu$. (b) Assortative mating ($\mu=0.5$) The $m$-carriers mate randomly, and $M$-carriers practise assortative mating with a probability of $\mu$ and mate randomly with a probability of $1-\mu$. On the vertical axes in (a) and (b), we also display the equilibrium values in the absence of migration. The situation in niche 2 leads to equivalent results, because the niche conditions are symmetric.}
\end{figure}

\begin{figure}
% ~/Forschung/MatlabFiles/Scritto/BreidbachJostSchindler/makeFigure_auswertungSims_timeToEqu.m
\subfigure[]{\includegraphics[width=7cm]{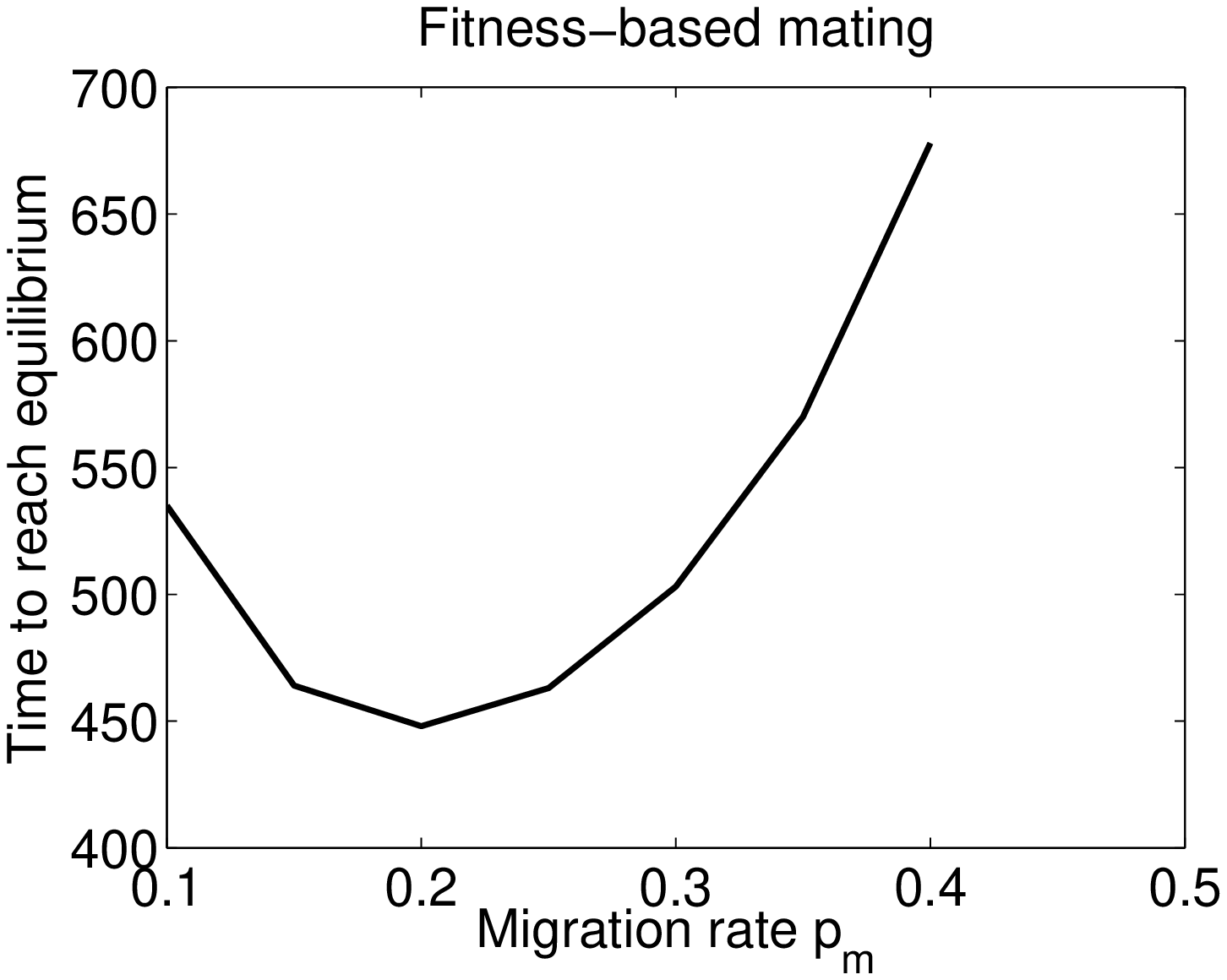}}
\subfigure[]{\includegraphics[width=7cm]{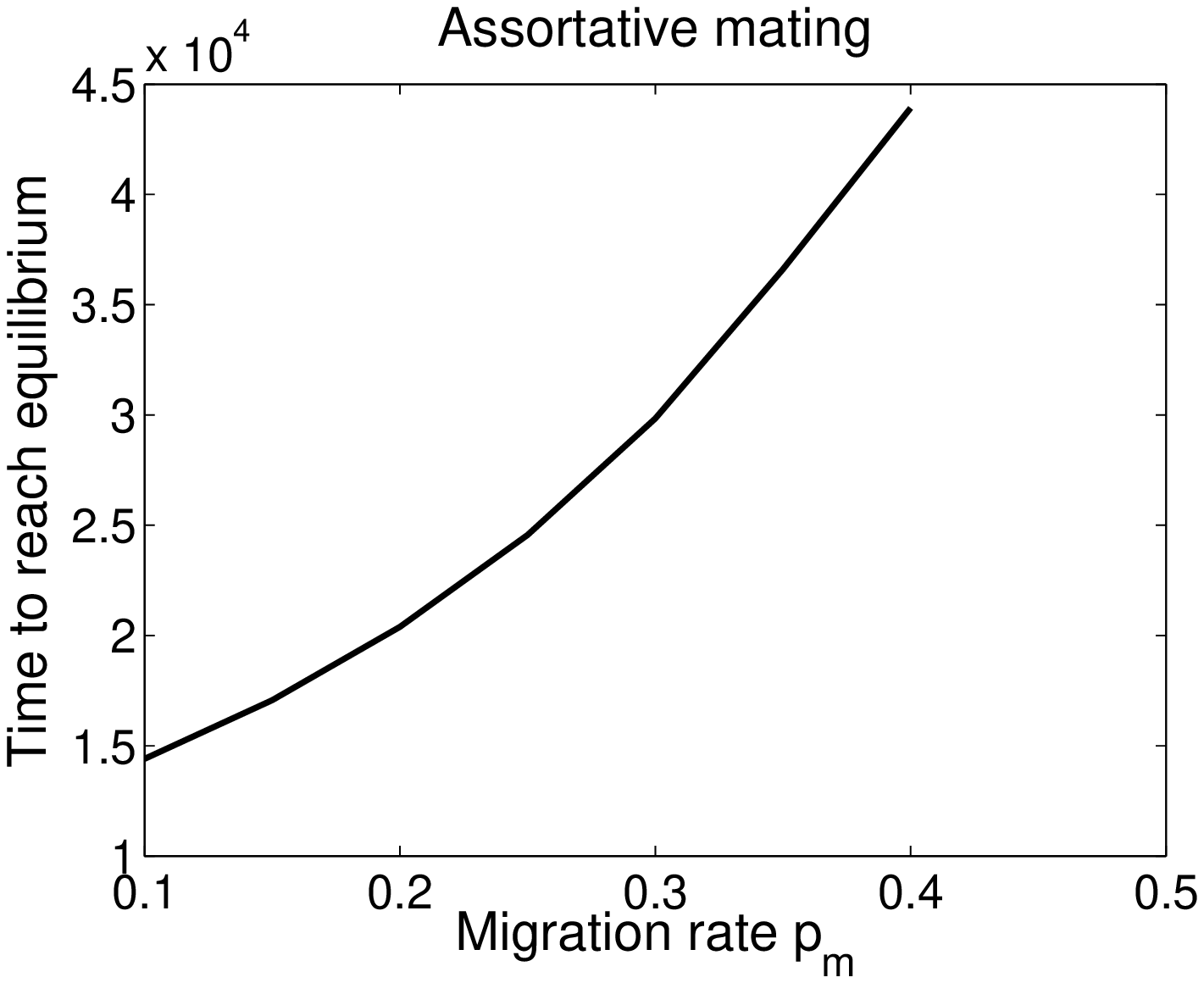}}
\caption{\label{fig:Time}The parameters and the initial gamete distribution have been set as in figure \ref{fig:zeitlicheEvo}. The plots show the point in time at which the equilibrium population is reached for various migration rates. (a) Fitmating ($\mu=0.5$) The $m$-carriers mate randomly, and $M$-carriers practise fitmating with a probability of $\mu$ and mate randomly with a probability of $1-\mu$. We see in that there is an intermediate migration rate for which the population reaches fastest the equilibrium. (b) Assortative mating ($\mu=0.5$) The $m$-carriers mate randomly, and $M$-carriers practise fitmating with a probability of $\mu$ and mate randomly with a probability of $1-\mu$.}
\end{figure}

In the simulations the population dynamics reaches an
equilibrium with the following properties. 
\begin{enumerate}
\item The allele $M$ for fitmating becomes fixed in a wide
  range of initial populations (Exceptions are when one of the alleles is already initially absent in both niches. Clearly, this allele would not reappear due to lacking mutation). In particular, the allele $M$ establishes
  itself faster than an allele for assortative mating would under
  otherwise equal circumstances. At figure \ref{fig:zeitlicheEvo}, for instance, we see that fitmating becomes fixed within the first 500 generations, whereas assortative mating needs 20000 generations until fixation.
\item Fitness-based mating leads to a higher equilibrium frequency
   of
  the fitter allele in each subpopulation. Thus, fitmating leads to a
  stronger divergence between the subpopulations than random or even
  assortative mating,
  thereby possibly enhancing an incipient speciation process. At figure \ref{fig:MigRate}, for instance, we see that fitmating, i.e., the $M$-allele, goes to fixation for all migration rates $p_m\in(0,\frac{1}{2})$, whereas assortative mating cannot outcompete random mating if migration is high, i.e., $p_m>0.3$.
\item A higher migration rate can speed up the approach to equilibrium
  as it increases the selective pressure in favor of the fitness-based
  mating allele $M$ which then in turn increases the selective
  advantage of the fitter allele in each subpopulation. At figure
  \ref{fig:Time}, for instance, we see that the point in time at which
  the equilibrium is attained under fitmating initially decreases,
  i.e., when $p_m\in(0,0.2)$, whereas under assortative mating, the
  population needs increasingly a longer time when $p_m$
  increases. Beyond $p_m=0.2$, the time of approaching the equilibrium
  increases with $p_m$ in both models, but it is still
  considerable lower than in the assortative mating model.
\item The $M$-allele can become dominant quite rapidly, already after
  a few hundred generations, as seen again in figures
  \ref{fig:zeitlicheEvo} and 
  \ref{fig:Time}.
\end{enumerate}
In particular,  at the final equilibrium when all individuals exhibit fitmating, the fitmating may not be
distinguishable from assortative mating because then in each niche,
 only the fitter males are chosen. They are chosen by fit resident females as well as by a modicum of less fit females that have either immigrated or resulted from ``mixed'' matings. So the major part of matings take place between partners with the same first locus allele. When we restrict in our model the migration to males
only, the equilibrium and the equivalence of fitness-based and assortative
mating will be reached even faster. Our point, however, is that during
the transient process, the two mechanisms of fitness-based and
assortative mating are clearly distinct, and the former produces
superior results. 


% \subsubsection{Fitness-based mating can induce assortative mating}

% In the model presented, each individual gets its turn in
% reproduction. This is an unrealistic assumption which allows for more
% genetic mixing resulting in more homogenous gene pools than is to be
% expected in a scenario where the overall number of matings is
% restricted by limitations of individual mating numbers. In more
% realistic conditions, the number of matings is usually bounded by time
% and energy constraints. Hence, individuals would not remain in the
% mating pool or would not return more often than a certain number of
% times. Assuming only one mating, i.e. monogamy, fit individuals would
% disappear from  the mating pool faster than less fit
% individuals. Therefore the number of matings between differently fit
% individuals will be lower, because fit ones choose other fit
% individuals. As a consequence, fit individuals mate assortatively and less fit individuals have to choose an equally less fit mating partner. In this way, fitness-based mating can induce assortative mating.

\section{Summary}

Inspired by many biological observations of preferences for fit mating
partners, in particular among females accepting only the fittest
males, typically leading to fierce competitions among males, we have developed
and 
implemented a model allowing for such a preference of fit mating
partners, called fitness-based mating.  We have shown that fitness-based mating is superior to random or even
 assortative mating in enhancing genetic differences between
 subpopulations in environments with different selective pressures. Fitness-based mating amplifies
natural selection because fit males then heap the benefit of better
access to females upon their own fitness advantage, and even less fit
females profit from the opportunity of producing fitter offspring with
fitter mates. When we then consider subpopulations occupying several 
niches with different selective pressures so that different alleles
induce higher fitness in the different niches, fitness-based mating
leads to a higher selection-migration equilibrium value in the niches,
as quantified by the frequency of the fittest genotype. Thus,
fitness-based mating can induce stronger polymorphism than random mating
 by maintaining a higher equilibrium frequency for well adapted
 genotypes in the different niches. Such a polymorphism could then
 trigger further divergence resulting in reproductive isolation, thus,
 speciation. It seems that fitness-based mating could easily induce
 runaway sexual selection as there will be a strong pressure for
 producing or even faking the phenotypic trait that indicates genetic
 fitness, but this issue is not explored in the present paper as it is
 already amply discussed in the literature and since it might distract from
 our main point which is at a more basic level. 

\bibliographystyle{abbrv}
\bibliography{biblevo11-01-21}

\end{document}